\documentclass[%
reprint,
 superscriptaddress,
 amsmath,amssymb,
 aps,
 prl,
]{revtex4-1}
\usepackage{graphicx}
\usepackage{dcolumn}
\usepackage{bm}

\begin{document}


\title{Physical white chaos generation}
\author{AnBang Wang}
\author{YunCai Wang}%
 \email{wangyc@tyut.edu.cn}
\author{BingJie Wang}
\author{Lei Li}
\author{MingJiang Zhang}
\affiliation{%
 Key Laboratory of Advanced Transducers and Intelligent Control System (Taiyuan University of Technology), Ministry of Education and Shanxi Province, Taiyuan 030024, China
}%
\affiliation{%
 Institute of Optoelectronic Engineering, College of Physics and Optoelectronics, Taiyuan University of Technology, Taiyuan 030024, China
}%
\author{WenDong Zhang}
\affiliation{%
 Key Laboratory of Advanced Transducers and Intelligent Control System (Taiyuan University of Technology), Ministry of Education and Shanxi Province, Taiyuan 030024, China
}%
\affiliation{%
 MicroNano System Research Center, College of Information Engineering, Taiyuan University of Technology, Taiyuan 030024, China
}%

\date{\today}

\begin{abstract}
Physical chaos is a fascinating prospect for high-speed data security by serving as a masking carrier or a key source, but suffers from a colored spectrum that divulges system's intrinsic oscillations and degrades randomness. Here, we demonstrate that physical chaos with a white spectrum can be achieved by the optical heterodyning of two delayed-feedback lasers. A white chaotic spectrum with 1-dB fluctuation in a band of 11 GHz is experimentally obtained. The white chaos also has a perfect delta-like autocorrelation function and a high dimensionality of greater than 10, which makes chaos reconstruction extremely difficult and thus improves security.
\begin{description}
\item[PACS numbers]
42.65.Sf, 05.45.Jn
\end{description}
\end{abstract}


\maketitle

Macroscopic chaos in fast nonlinear optoelectronic or laser systems has motivated remarkable developments in information security. For instance, macroscopic chaos serving directly as a masking carrier improves the rate of secure communications to gigabits per second \cite{VanWiggeren1998, *Argyris2005}. Additionally, fast chaos replaces microscopic noise as an entropy source to greatly increase the speed of generating physical random keys \cite{Uchida2008, *Yoshimura2012, Reidler2009, *Kanter2010} and as a radar signal to enhance anti-jamming capability and spatial resolution \cite{Lin2004l, *Lin2004r, YWang2008, *AWang2012}. One of the most important factors is that the physical chaos has a spread spectrum and an irregular waveform with larger amplitude than that of microscopic noise, and the use of chaos can avoid the wideband amplification required in a noise generator.

Essentially, the randomness and security of chaos is vital to chaos-based information security. However, the randomness of physical chaos is usually degraded by intrinsic oscillations of nonlinear chaos systems. Currently, fast physical chaos generators typically consist of a nonlinear optical or optoelectronic device subject to a feedback loop that partially couples the output back into the device, such as an external-cavity semiconductor laser (ECL) \cite{Albert2011, *Fischer1994} and an optoelectronic oscillator \cite{Lavrov2009, *Callan2010, *Nourine2011}. As a result, the intrinsic oscillations, including the characteristic response frequency of the nonlinear device and the resonance of the feedback loop, introduce deterministic features and periodicity into the chaos and thus limit the randomness. Taking the paradigm of a delay system, i.e., an ECL, as an example, the relaxation oscillation of the semiconductor laser dominates the laser intensity chaos and leads to a strong peak projecting from the power spectrum \cite{AWang2008}. Additionally, the resonance of the external feedback cavity excites many external-cavity modes (ECMs) in the laser field and leads to a periodically modulated spectrum similar to that of a frequency comb with a spacing equal to the reciprocal of the feedback delay \cite{Rontani2009, Hirano2009}. Considerable effort has recently been devoted to expand the power spectrum \cite{AWang2008, Takiguchi2003, *Uchida2003, *Zhang2011, *AWang2013APL} or to depress the external-cavity resonances \cite{Rontani2007, *Wu2009, *SLi2012}, but it is difficult to completely eliminate the effects of the intrinsic oscillations \cite{Reidler2009, *Kanter2010, YWu2012}. These intrinsic oscillations impose restrictions on random number generation from physical chaos \cite{Hirano2009} and raise arguments that the extracted numbers are not truly random \cite{Guo2010}. Worse yet, the intrinsic oscillations are easily intercepted from time series \cite{YWu2012, Rontani2009} and then pose a risk of cracking to chaos systems \cite{Hegger1998, *Bunner1996}.

A reliable physical chaos should have a wide and flat power spectrum containing various frequency components with homogeneous amplitudes so that system's intrinsic frequencies cannot be determined. Such a chaos having a white spectrum is referred to as white chaos, which has only been founded in some mathematical maps such as the piecewise-linear Markov map \cite{Kohda1992} and the Logistic map \cite{McGonigal1987}. However, the spectral width of the circuit realization of the mathematical chaos is extremely limited at several kilohertz \cite{McGonigal1987}, which is much lower than the present information speed.

In this Letter, we present a scheme that can generate broadband white chaos by using the optical heterodyning of two ECLs. The heterodyning convert laser phase dynamics into intensity that can eliminate the feature of laser relaxation oscillation \cite{AWang2013}. Simultaneously, the heterodyning can conceal both external-cavity signatures of two ECLs in the case of external cavities with different lengths.

The schematic of the proposed method is outlined in Fig.1.
\begin{figure}[t]
\includegraphics[width=3.375in]{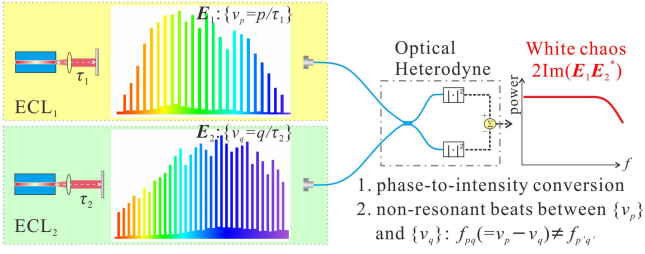}
\caption{\label{fig:one} (color online) Schematic and principle. Each ECL has a widened comb-like optical spectrum including ECMs, i.e., $\nu_{p}=p/\tau_{1}$, $\nu_{q}=q/\tau_{2}$. Optical heterodyning can convert laser phase dynamics into intensity, yielding a wideband spectrum without the signature of relaxation oscillation, and induce non-resonant beats between $\nu_{p}$ and $\nu_{q}$ when $\tau_{1}$ and $\tau_{2}$ are incommensurate, eliminating the signature of external-cavity resonances.}
\end{figure}
Because of external-cavity feedback, each ECL has a widened optical spectrum including comb-like ECMs, i.e., $\nu_{p}=p/\tau_{1}$ and $\nu_{q}=q/\tau_{2}$, where $p$ and $q$ are positive integers. The widened optical spectrum indicates fast laser phase dynamics. The optical heterodyning of the two ECLs can convert the fast phase dynamics into intensity, and then achieve a flat beat spectrum by properly tuning the lasers' frequencies. Furthermore, if the feedback delays $\tau_{1}$ and $\tau_{2}$ are incommensurate ($q\tau_{1} \neq p\tau_{2}$), the difference frequencies between the two sets of ECMs are totally non-resonant, i.e., $f_{pq} (\equiv p/\tau_{1}-q/\tau_{2}) \neq f_{p'q'}$. Thus, the heterodyning cannot produce any dominant component in the beat signal, and the beat spectrum therefore can be free of periodic modulation related to ECMs. Therefore, in principle, the heterodyne signal can have a white spectrum.

Theoretically, the beat signal is expressed as $I_{bt}= 2A_{1}(t)A_{2}(t)\sin[2\pi\Delta\nu_{0}t+\varphi_{1}(t)-\varphi_{2}(t)]$, where $A_{i}(t)$, $\varphi_{i}(t)$, and $\nu_{0i}$ $(i=1, 2)$ are the amplitude, phase, and solitary frequencies of the two lasers, respectively, and $\Delta\nu_{0} = \nu_{01}-\nu_{02}$ is the frequency difference. The generation of white chaos can therefore be proved numerically by using the following Lang-Kobayashi equations \cite{LK1980}.
\begin{eqnarray}
&\dot{\textit{A}}=\frac{1}{2}[\textit{G}-\tau_p^{-1}]\textit{A}+\tau_{in}^{-1}\rho\textit{A}(t-\tau)\cos\theta,\\
&\dot{\varphi}=\frac{1}{2}\alpha[\textit{G}-\tau_p^{-1}]-\tau_{in}^{-1}\rho[\textit{A}(t-\tau)/\textit{A}]\sin\theta,\\
&\dot{\textit{N}}=\textit{J}-\tau_N^{-1}\textit{N}-\textit{G}\textit{A}^2,
\end{eqnarray}
where, $N$ is carrier density, $\theta=2\pi\nu_{0}\tau+\varphi-\varphi(t-\tau)$ and $G=g(N-N_{0})/(1+\varepsilon A^2)$, $\kappa=10\log_{10}(\rho^2)$ represents the feedback strength, $\tau$ is the feedback delay. In simulations, the transparency carrier density $N_0=0.455\times10^6 \mu \textnormal{m}^{-3}$, the differential gain $g=1.414\times10^3 \mu \textnormal{m}^{3}\textnormal{ns}^{-1}$, the gain saturation parameter $\varepsilon=5\times10^{-5} \mu \textnormal{m}^{3}$, the carrier lifetime $\tau_{N}=2.5$ ns, the photon lifetime $\tau_{p}=1.17$ ps, the linewidth enhancement factor $\alpha=5.0$, the round-trip time in laser cavity $\tau_{in}=7.38$ ps, the threshold current density $J_{th}=4.239\times10^{5} \mu \textnormal{m}^{-3}\textnormal{ns}^{-1}$, and the bias current $J=1.7J_{th}$, which yields a relaxation frequency of approximately 3 GHz.

Figure 2a shows the optical spectra of two ECLs, which were numerically obtained with $\tau_{1}=1.0$ ns, $\tau_{2}=2.77$ ns, feedback strength $\kappa_{1,2}=-13.98$ dB, $\nu_{01}=193.548$ THz and $\nu_{02}=193.558$ THz.
\begin{figure}[b]
\includegraphics[width=3.375in]{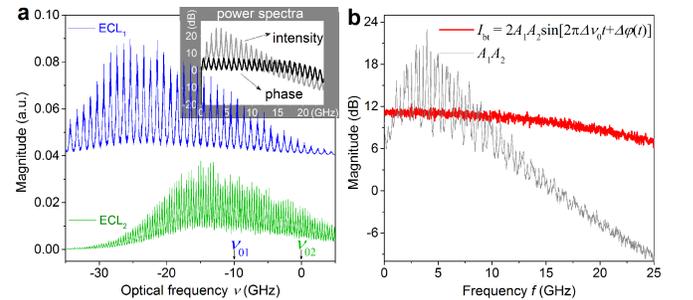}
\caption{\label{fig:two} (color online) A numerical demonstration of white chaos. \textbf{a}, Optical spectra of two ECLs. The inset plots the power spectra of the laser intensity and phase $\sin\varphi$ of $\textnormal{ECL}_{1}$. \textbf{b}, Power spectra of the beat signal (red) and $A_{1}A_{2}$ (gray). $\tau_{1}=1.0$ ns, $\tau_{2}=2.77$ ns, $\kappa_{1,2}=-13.98$ dB, $I_{1,2}=1.7I_{th}$, $\nu_{01}=193.548$ THz and $\nu_{02}=193.558$ THz.}
\end{figure}
Both optical spectra are red-shifting, cover a broad band of approximately tens of gigahertz, and have comb-like modes. The general power spectral properties of laser intensity and phase are depicted by those of $\textnormal{ECL}_{1}$ in the inset. Shown as the thin line, the spectrum of the laser intensity not only has a dominant peak at approximately the relaxation frequency but also has a pattern of periodic modulation, which is due to resonant difference frequencies $f_{pp'} =(p-p')/\tau_{1}$ of ECMs. The modulation depth is larger than 10 dB around the relaxation frequency and increases further with the feedback strength. As shown as the thick line, the spectrum of the laser phase $\sin[\varphi(t)]$ also has the signature of periodic modulation but with a smaller depth. Interestingly, its overall appearance is flat in a range of 0-15 GHz and does not have a dominant component such as the relaxation oscillation. A possible explanation is that the effect of optical feedback on the phase dynamics exceeds that of the carrier change as the feedback strength increases, and thus the relaxation oscillation in phase signal is no longer dominant. These spectral properties also indicate that the phase dynamics is different from laser phase noise because the latter is narrow band and free of any external-cavity resonance \cite{Guo2010}.

The power spectrum of the numerical beat signal is shown as the red line in Fig.~\ref{fig:two}b. Clearly, the spectrum is white in a broad frequency range of approximately 0-12 GHz; it no longer has dominant frequency components related to relaxation oscillation, nor does it have periodic modulation related to external-cavity resonance. By comparison, the spectrum of $A_{1}A_{2}$ (i.e., the sinusoidal term related to beating is removed from $I_{bt}$) shown as the gray line is similar to that of laser intensity, which clearly shows the contribution of the optical heterodyning to the generation of the white chaos.

We mention that the concealment of the periodic spectral pattern can be understood from the viewpoint of autocorrelation. The periodic spectral pattern means that the chaos has correlation at the delay value and its integer multiples \cite{Rontani2009, Hirano2009, YWu2012}. Because the two chaotic lasers are independent, the autocorrelation function (ACF) of $I_{bt}$ is not larger than the product of the ACFs of laser amplitudes $A_{1, 2}$. The product therefore mutually conceals the correlation at the delay values under the condition $q\tau_{1} \neq p\tau_{2}$ meaning $\textnormal{ACF}_{1}(p\tau_{2})= \textnormal{ACF}(q\tau_{1})=0$. In addition, because the number of correlation peaks of laser chaos is finite, the constraint can be relaxed: integers $p$ and $q$ can be commensurate, but they must be large enough, for instance $\tau_{1}/\tau_{2}=100/277$ in the numerical demonstration, to satisfy that the laser chaos does not have correlation at $q\tau_{1}$ or $p\tau_{2}$, i.e., $\textnormal{ACF}_{1,2}(q\tau_{1}=p\tau_{2})=0$.

In our experiments, each ECL consists of a distributed feedback semiconductor laser subject to optical feedback from a fibre mirror. The two feedback delays are $\tau_{1}=85.3$ ns and $\tau_{2}=110.7$ ns, respectively. In each external cavity an optical variable attenuator is used to adjust the feedback strength and then change the chaotic state. The lasers' frequencies are tuned slightly by using high-precision temperature controllers to adjust the operating temperature. The extraction of the beat signal between the two lasers is implemented by a fast balanced photodetector. Note that the two fiber paths from the coupler to the photodetector should have same transmitting length and loss. (See system details in Section S1 in \cite{Suppl})

We now demonstrate a white chaos that was experimentally generated with parameters: $I_{1}=14.8$ mA ($1.44I_{th1}$), $\kappa_{1}=-15.9$ dB, $I_{2}=14.8$ mA ($1.35I_{th2}$), $\kappa_{2}=-19.5$ dB and $\Delta\nu=-12.5$ GHz (wavelengths $\lambda_{1}=1549.980$ nm, $\lambda_{2}=1549.882$ nm). Note that the optical spectra of the two chaotic lasers are plotted in the inset of Fig. 3a, and we use their center frequencies to calculate the frequency difference due to the red-shifting.
\begin{figure}[t]
\includegraphics[width=3.375in]{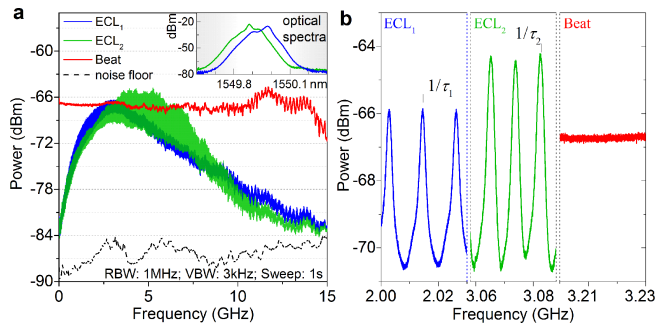}
\caption{\label{fig:three} (color online) Experimental power spectra in different frequency scales: \textbf{a}, 15 GHz, and \textbf{b}, 30 MHz (red, the white chaos; blue, $\textnormal{ECL}_{1}$; green, $\textnormal{ECL}_{2}$). $I_{1}=1.44I_{th1}$, $\tau_{1}=85.3$ ns, $\kappa_{1}=-15.9$ dB, $I_{2}=1.35I_{th2}$, $\tau_{2}=110.7$ ns, $\kappa_{2}=-19.5$ dB, $\Delta\nu=-12.5$ GHz (the inset plots the lasers' optical spectra).}
\end{figure}
Figure~\ref{fig:three}a shows the power spectrum of the beat signal as the red line as well as the spectra of the intensity chaos of $\textnormal{ECL}_{1}$ and $\textnormal{ECL}_{2}$ as the blue and green lines, respectively, in a large scale of 15 GHz. The spectrum of each ECL has an obvious peak around the laser relaxation frequency, which is approximately 18 dB higher than the lowest component. By contrast, the beat spectrum is very flat in a wide band and has no dominant peak. Specifically, the spectral fluctuation is only $\pm1.0$ dB in the band from 0 to 11 GHz and is $\pm1.5$ dB in the band from 0 to 14.3 GHz. It is worth noting that the $-3$-dB bandwidth of 14.3 GHz exceeds that of some reported amplified spontaneous emission noises \cite{XLi2011, *Argyris2012, *Williams2010}. Furthermore, as depicted in the left and middle panes in Fig.~\ref{fig:three}b, the fine spectra in a scale of 30 MHz of the laser chaos have a pattern of periodic modulation with a depth of greater than 5 dB. Note that the fine spectrum will exhibit multiple periods for feedback from multi-cavities \cite{YWu2012}. In contrast, as shown in the right pane, the fine spectrum of the beat is straight, having a slight random fluctuation with a standard deviation of 0.06 dB. Therefore, both the entire and fine spectra show that the beat signal is a broadband white chaos.

Figure 4a shows the time series and the corresponding probability distribution function of the beat signal as well as the intensity chaos of ECLs, which were recorded by using a real-time oscilloscope with an analogue bandwidth of 6 GHz.
\begin{figure}[t]
\includegraphics[width=3.375in]{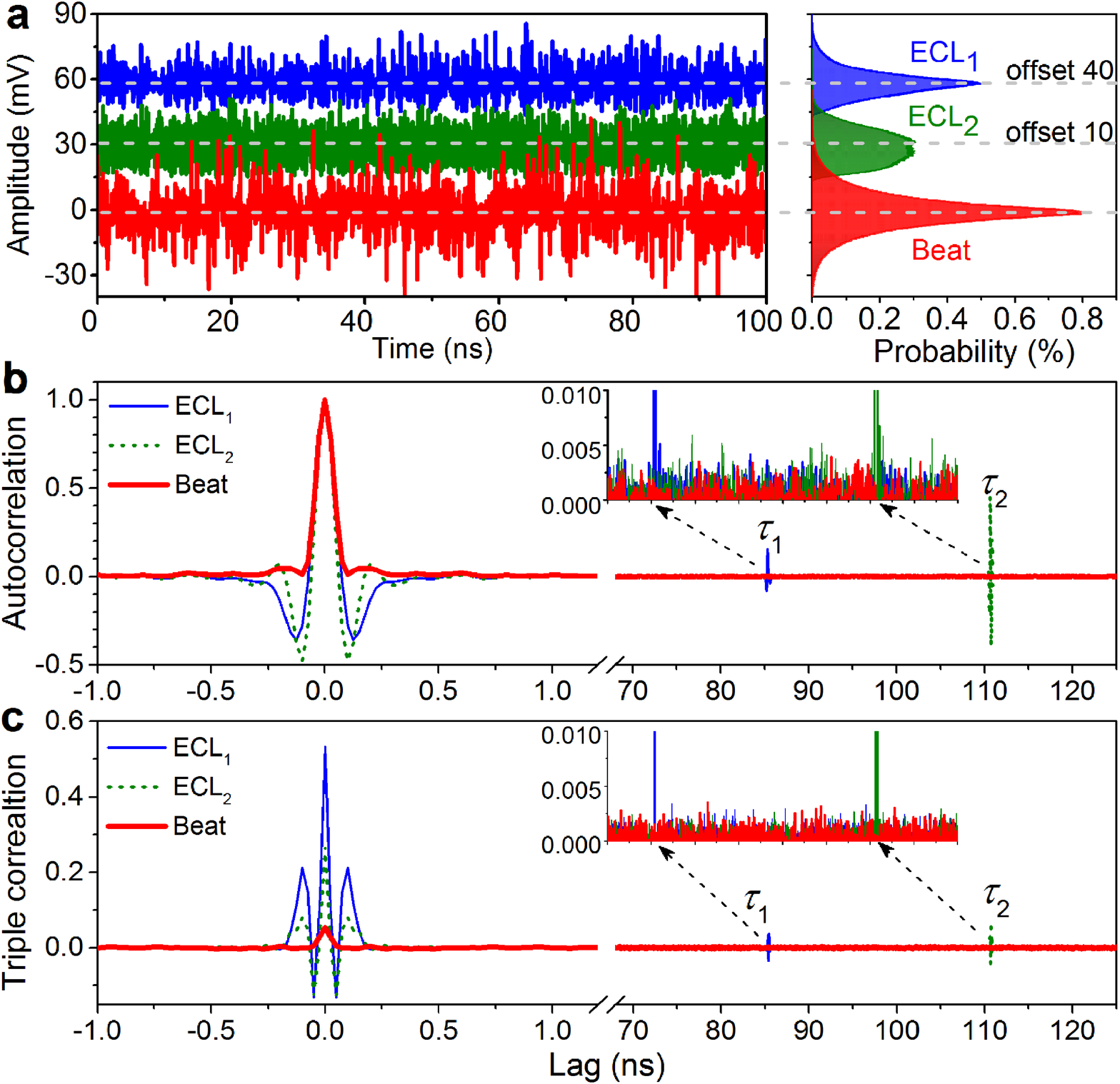}
\caption{\label{fig:four} (color online) Temporal properties of experimental white chaos. \textbf{a}, Time series (left) and amplitude probability distributions (right), \textbf{b}, autocorrelation functions, and \textbf{c}, triple autocorrelation functions (red, the white chaos; blue, $\textnormal{ECL}_{1}$; green, $\textnormal{ECL}_{2}$). Data length is 2M points for calculating correlation functions. The experimental conditions are the same as in Fig. 3.}
\end{figure}
Clearly, the amplitude distribution of the beat signal is more symmetrical; the skewness of the beat signal is 0.054, which is significantly less than the skewness values of 0.533 and 0.267 of the intensity chaos of $\textnormal{ECL}_{1}$ and $\textnormal{ECL}_{2}$, respectively. Because of the distribution symmetry of the beat signal, nearly equal numbers of ones and zeros can be achieved in random number generation \cite{Reidler2009, *Kanter2010, Argyris2012, *Williams2010}. We mention that the amplitude distribution of the beat signal exhibits a Lorentzian profile, meaning the beat is different from the Gaussian spontaneous emission noise characteristic of lasers.

Figure~\ref{fig:four}b shows the ACFs of the beat and the laser chaos, which are calculated as $E[(f(t)-\mu)(f(t- \tau)- \mu)]/\sigma^{2}$, where $E[\cdot]$ represents the expected value operator, and $\mu$ and $\sigma$ are the mean value and standard deviation of a signal $f(t)$, respectively. As shown as the blue and green lines in Fig.~\ref{fig:four}b, the autocorrelation trace of the laser chaos exhibits an obvious oscillation after the main peak at zero lag, which results from the dominant relaxation oscillation. Additionally, the trace has a distinct peak at the feedback delay, i.e., $\textnormal{ACF}_{1}(\tau_{1})= 0.155$ and $\textnormal{ACF}_{2}(\tau_{2})= 0.416$, corresponding to the spectral modulation period shown in Fig.~\ref{fig:three}b. By contrast, the autocorrelation trace of the beat signal shown as the red line has only one peak located at zero lag without obvious oscillation. Significantly, the correlation values at lags $\tau_{1}$ and $\tau_{2}$ are $-2.5\times10^{-4}$ and $1.2\times10^{-4}$, respectively, which are less than the standard deviation of $1.14\times10^{-3}$ of the background noise. The beat signal therefore has a ACF profile like a Dirac delta function.

Furthermore, the higher-order autocorrelation of the beat signal also exhibits properties similar to those of white noise. As an example, we take the triple autocorrelation, which is defined as $E[(f(t)-\mu)(f(t-\tau)- \mu)(f(t-\tau')- \mu)]/\sigma^{3}$. The triple autocorrelation is a two-dimensional function (see Section S2 in \cite{Suppl}), and for the sake of brevity we show in Fig.~\ref{fig:four}c the sliced traces at $\tau'=2\tau$. As plotted as the blue and green lines, the triple autocorrelation trace of the laser chaos still has a feature of the laser relaxation oscillation and a correlation peak at the feedback delay. Depicted as the red line, the triple correlation of the beat signal does not have these features. It has only a small peak at zero lag and is approximately zero for other values of lag. Such a triple correlation indicates that the beat signal has a flat higher-order spectrum (see Section S3 in \cite{Suppl}). We notice that the peak value at zero lag is the skewness of signal's amplitude distribution mentioned above.

We further utilize correlation dimensional analysis to show that the generated white chaos is high-dimensional by using the Grassberger-Procaccia (GP) algorithm \cite{GP1983}. The algorithm counts the correlation integral $C(r)$ that is the probability of pairs of points with a mutual distance not larger than $r$ in a delay-embedded phase space and estimates the correlation dimension by the converged slope of logarithmic plots of $C(r)$ versus $r$. For comparison we have analyzed the laser chaos of $\textnormal{ECL}_{1}$ and the white chaos obtained both in simulations and experiments. Figure 5a plots the results for the numerical laser chaos with a data length $N = 30,000$ points.
\begin{figure}[t]
\includegraphics[width=3.375in]{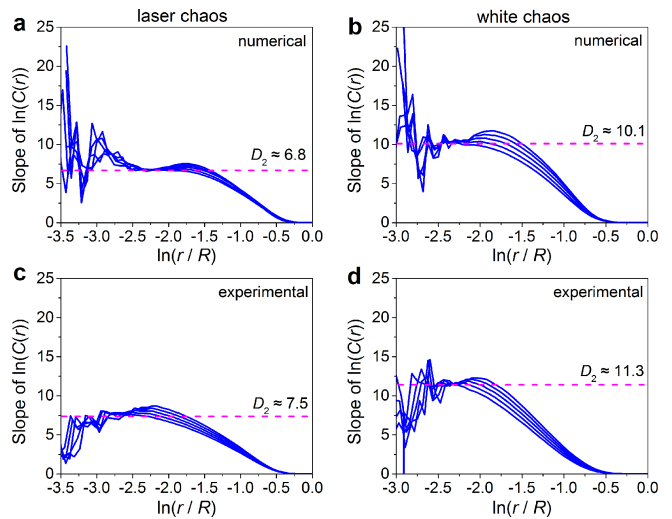}
\caption{\label{fig:five} (color online) Correlation dimension analysis for the laser chaos (\textbf{a}, numerical; \textbf{c}, experimental) and the white chaos (\textbf{b}, numerical; \textbf{d}, experimental) by the slope of the logarithmic plot of the correlation integral $C(r)$. The horizontal coordinates are normalized by the diameter $R$ of the reconstructed attractor.}
\end{figure}
A clear scaling region of $\ln(r/R)$ is recognized between $-2.5$ and $-2.1$, yielding a dimension $D_{2} \simeq 6.8$, where $R$ is the diameter of the reconstructed attractor. Similar clear convergence can also be found for the laser phase dynamics (see Section S4 in \cite{Suppl}). Because the white chaos has a flat and wide power spectrum and may thus have higher dimensionality, we increase the data length to 50,000 points to attempt estimation for the white chaos. As depicted in Fig.~\ref{fig:five}b, a plateau still appears but in a smaller scaling region from $-2.3$ to $-2.1$. The plateau indicates the dimension of the numerical white chaos as approximately 10.1, which is very close to the estimation bound of 10.3 given by $2\ln(N)/\ln(R/r)$ in the scaling region \cite{Eckmann1992}. For the experimental chaos, to reduce the influence of detection noise, we have employed a re-embedding method that applies the GP algorithm on the principal components obtained by singular value decomposition on time series \cite{Fraedrich1993}. The results of the experimental laser chaos and white chaos, which are plotted in Figs.~\ref{fig:five}c and ~\ref{fig:five}d, respectively, also show that the white chaos has higher dimensionality than the laser chaos. We mention that for the white chaos the correlation dimension of 11.3 actually exceeds the estimation bound, implying a higher dimensionality. Accurate estimation of the dimension for the experimental white chaos requires a huge dataset and a time-consuming re-embedding method, and thus is difficult. It is therefore believed that the high-dimensional white chaos can avoid possible phase-space reconstruction.

At last, we discuss that the two lasers should be in states of developed chaos under moderate or strong feedback, so that they have fast phase dynamics to achieve white chaos. Otherwise, even if one of the lasers is in an undeveloped chaos state under weak feedback, the beat spectrum will not be smooth and will have a strong peak at the frequency equal to the optical frequency difference (see Section S5 in \cite{Suppl}). But in this case the spectral periodic pattern is still canceled due to the existence of the non-resonant beating effect of ECMs. If one laser becomes solitary by removing feedback, the beat spectrum will be similar to the optical spectrum of another ECL and inherit the periodic modulation feature \cite{Brunner2012}. In addition, the optical frequency difference between the two lasers should be adjusted according to the lasers' optical spectral widths. If the two laser spectra are separate or coincide, the beat spectrum exhibits an inclined profile with a strong center frequency equal to the optical frequency difference (Section S5 in \cite{Suppl}). A zero frequency difference will yield an unsmooth spectrum similar to that of the delayed self-interference of a chaotic laser \cite{AWang2013}.

To sum up, we demonstrate a simple scheme that uses the optical heterodyning of two ECLs to generate physical white chaos. Experiments and simulations show that the generated white chaos is high-dimensional and broadband. The defects in laser chaos such as the colored spectrum and the asymmetric distribution, are removed. Thus, the white chaos not only has the merit of large amplitude but also gains randomness similar to white noise. Furthermore, the elimination of the system features and the high dimensionality make the white chaos more reliable for applications to data security because possible phase-space reconstruction can be avoided. In addition, our results will excite research interest in the generation of white noise using physical chaos \cite{Kautz1999}.

The authors thank Prof. I. Fischer for his helpful suggestions. This work was supported by NSFC (Nos. 60908014, 61108027, 61205142, and 61227016), by the International Science and Technology Cooperation Program of China (No. 2014DFA50870), and by the Shanxi Innovative Research Team for Key Science and Technology, China.

\nocite{*}


\providecommand{\noopsort}[1]{}\providecommand{\singleletter}[1]{#1}%

\end{document}